\documentclass[aps,prl,twocolumn,superscriptaddress,showpacs,a4paper]{revtex4}

\usepackage{graphicx,amsmath,amssymb}

\begin{document}
\title{Density of states governs light scattering in photonic crystals}

\author{P.D. Garc\'{i}a}
\affiliation{Instituto de Ciencia de Materiales de Madrid (CSIC)
 and Unidad Asociada CSIC-UVigo, Cantoblanco 28049 Madrid
Espa\~{n}a.}

\author{R. Sapienza}
\affiliation{Instituto de Ciencia de Materiales de Madrid (CSIC)
 and Unidad Asociada CSIC-UVigo, Cantoblanco 28049 Madrid
Espa\~{n}a.} \homepage[]{www.icmm.csic.es/cefe/}
\email[]{sapienza@icmm.csic.es}

\author{Luis S. Froufe-P\'{e}rez}
\affiliation{Instituto de Ciencia de Materiales de Madrid (CSIC)
 and Unidad Asociada CSIC-UVigo, Cantoblanco 28049 Madrid
Espa\~{n}a.}

\author{C. L\'{o}pez}
\affiliation{Instituto de Ciencia de Materiales de Madrid (CSIC)
 and Unidad Asociada CSIC-UVigo, Cantoblanco 28049 Madrid
Espa\~{n}a.}

\date{\today}

\begin{abstract}
We describe a smooth transition from (fully ordered) photonic
crystal to (fully disordered) photonic glass that enables us to
make an accurate measurement of the scattering mean free path in
nanostructured media and, in turn, establishes the dominant role
of the density of states. We have found one order of magnitude
chromatic variation in the scattering mean free path in photonic
crystals for just $\sim 3\%$ shift around the band-gap ($\sim 27$
nm in wavelength).
\end{abstract}


\maketitle Contemporary photonic science is capable of addressing
fundamental questions at the basis of light-matter interactions,
of which the role of the photon density of states,
$\mathcal{D}(\omega)$,  in nanostructured media is one of the most
intriguing.

Artificially engineered materials allow the control of light transport through interference in the
internal nanostructure, rather than on the refraction in the body
boundaries, engendering new materials properties.
Photonic crystals, in which the dielectric constant is
periodically modulated, manipulate electromagnetic states and the
available phase space and control fundamental aspects of
light-matter interaction like light emission \cite{yablo} and
light transport \cite{Saggio},  much like semiconductors control
electrons. Redistribution and inhibition of the emission from
photonic crystals was proven \cite{Lodhal}, but unconventional
light transport in partially-disordered photonic crystals has only
been hinted at by pioneering experiments \cite{Frank, Toninelli}.
Other topologies, like random media \cite{randommedia}, correlated
disordered \cite{PhGlAdMat} or fractal \cite{fractal}, employ the
aperiodic subwavelength dielectric nanostructure to achieve
similar light control for transport \cite{light diffusion} and
random lasing emission \cite{randomlasing}.

Light scattering by weak topological disorder in a photonic
crystal and the interplay between order and disorder has yet to be
fully understood and explored. As an important step, the relation
between scattering extintion and $\mathcal{D}(\omega)$ has just
been theoretically derived \cite{Carminati}.
As pointed out by John \cite{Saggio} a dramatic change in light
diffusion can occur for frequencies in or around the band-gap and
eventually Anderson localisation of light  can be reached, the
photonic conductor becoming an insulator \cite{Anderson}. In the
quest for light localization, the first experiments focused on
fully random media \cite{Diederik}, only recently transverse
localization has been reported in two-dimensional crystals with
disorder \cite{Schwartz}.

Even far from the localization regime, the scattering properties
of Bloch modes, the periodic electromagnetic modes of a photonic
crystal, are expected to be profoundly different from the
diffusive modes encountered in conventional random media.
Pioneering experiments on coherent backscattering
\cite{Foemius-cbs}, and diffuse light transport \cite{Astratov, Vlasov} in
photonic crystals searched for signatures of Bloch-mode
mediated scattering but have merely shown standard light
diffusion. Moreover, the experiments have been interpreted using a
model that assumes no photonic modal dispersion   but rather a
modified reflectivity at the system boundaries \cite{VosPRB}.

In this letter we study the scattering mean free path, $\ell_s$, the fundamental building block
for any wave transport model, for the special case of photonic crystals with a
controlled amount of disorder. We report experimental evidence of
strong chromatic dispersion of $\ell_s$ from band-edge to
band-gap, and values of up to $\sim 100-500$ $\mu$m, i.e.  $\sim
300$ times the lattice parameter $(a)$, an order of magnitude
higher than previously reported \cite{Foemius-cbs, Astratov, Vlasov}.


Single scattering events in a system with modified light modes and
density of states, as in a photonic crystal, are very different
from those occurring in vacuum due to: a) an increase of
light-matter interaction and thus of scattering by defects, when
$\mathcal{D}(\omega)$  is increased at the vicinity of band-edges
and b) a suppression of the scattering channels, i.e. a increase
of $\ell_s$ in the band-gap, where $\mathcal{D}(\omega)$ is
strongly reduced.

The scattering strength can be studied via simultaneous reflection
and transmission measurements, when absorption is negligible  (the
absorption length $\ell_a \sim 10$ m \cite{PRLmie}) and for
energies below the onset of diffraction ($a/\lambda \sim 1.12$
\cite{Floren} where $\lambda$ is the light wavelength).
We assume that scattering losses follow Lambert-Beer's law, i.e.  that after a
thickness $L$, a ballistic beam attenuates as $I(L)=I_0
\exp(-L/\ell_s)$.
The
intensity balance can then be expressed as
\begin{equation}
\label{lambertEq}
T(L)+R(L)=\exp(-L/\ell_s),
\end{equation}
where $T(L)$ and $R(L)$ are the ballistic transmission and
reflection as a function of sample thickness 
in a given direction.

\begin{figure}[!h]
    \begin{center}
   \includegraphics[width=6cm]{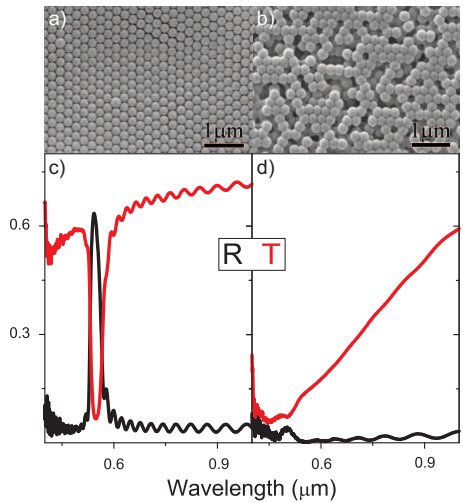}
    \caption{ \label{fig1}
    Top panel: SEM images of photonic crystal with 0\% (a) and 40\%  (b) vacancy doping.
    (c) and (d) reflection and transmission from the corresponding
    samples of the above panel. }
    \end{center}
    \end{figure}

The samples are Polymethil-metacrilate (PMMA, refractive index, $n
= 1.49$) self-assembled face centred cubic (fcc) photonic crystals
\cite{Colvin} with controlled density of intentionally added
vacancies \cite{PhGlAdMat} (from 0\% to 40\%) (see Fig.
\ref{fig1}a and \ref{fig1}b).
These vacancies are obtained upon removal of given fractions of
constituting spheres from random lattice positions.

The scattering mean free path in a random medium can be shown to
be $\ell_s = 1/\rho_s \sigma$, where $\sigma$ is the scattering
cross-section and $\rho_s$ is the scatterers number density
\cite{light diffusion}. $\ell_s$ is not only a measure of the
quality of a photonic structure but also the basic length-scale of
a more complex picture of multiple scattering and light diffusion.
While quantities like the transport mean free path or the
diffusion constant are meaningful only in the context of the
diffusion approximation, $\ell_s$ has a full meaning in any
microscopic picture, regardless of the transport regime.

In our system, the degree of extrinsic disorder (see scanning
electron microscopy (SEM) images in Fig. \ref{fig1}a and \ref{fig1}b) can be very
precisely and uniformly tuned, while keeping the sample thickness
controlled. This allows us to develop a setup to measure
Lambert-Beer's law for photonic crystals. We used a Fourier
transform spectrometer coupled to a microscope
allowing to probe the scattering in the (111) direction
while illuminating a
constant-thickness area with a spot of $\sim 80$ $\mu$m. The
spectra (see figures \ref{fig1}c and \ref{fig1}d) are taken in
adjacent regions
which are visible by optical microscope inspection as terraces on
the sample surface. The thickness of such layers is precisely
measured with an uncertainty of 2\%, via the density of
Fabry-Perot fringes that occur for the interference of the light
reflected from the front and rear faces of the sample.

   \begin{figure}[!b]
    \begin{center}
   \includegraphics[width=8.5cm]{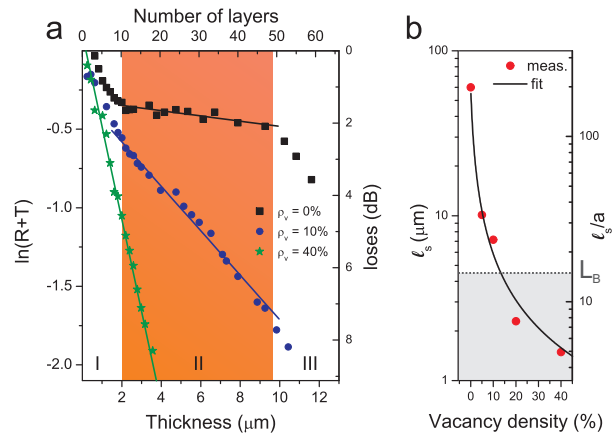}
    \caption{ \label{fig2}
    (a) Plot of $\ln (R+T)$ as a function of the sample thickness, at $\lambda =633$ nm,
     for different vacancy density doped photonic crystals (from 0\% to 40\% vacancies doping),
    of 237 nm diameter.  (b) $\ell_s$ obtained from linear fitting of the slope, for regime II.
    It also shows the Bragg length ($L_B$) in the case of $\rho_v = 0\%$ as shaded area. }
    \end{center}
    \end{figure}

Figure \ref{fig2}a shows the measured $\ln(T+R)$ for three
different degrees of vacancy doping \cite{PhGlAdMat} i.e. for
different degrees of extrinsic disorder, at a wavelength of 633 nm
and for spheres of 237 nm in diameter ($a/\lambda  = 0.52$). In
this type of representation, the slope yields directly
$(-\ell_s)^{-1}$ according to Eq. \ref{lambertEq}. This wavelength
is chosen to exemplify a spectral region where no photonic band
features are present, as, at such a low energy, the photonic
crystal band dispersion is the same as in a uniform homogeneous
effective medium. In figure \ref{fig2}a three scattering regimes
are clearly distinguishable. For thicknesses lower than
\textit{ca.} 10 layers, (regime I), up to $\sim 25-30\%$ of the
incident light is scattered due to surface effects, in the form of
stacking patterns \cite{Alvaro-monocapa}, high lattice
displacements and, even stacking order arrangements \cite{Denkov}.
When the second regime (II) sets in, the slope $\ell_s^{-1}$ reaches a
stationary value that characterises the photonic crystal. Eq. \ref{lambertEq}
holds and scattering losses scale with sample thickness like $\sim
\exp(L/\ell_s)$.
Finally, for larger thicknesses, a third scattering regime (III)
appears in Fig. \ref{fig2}a. Apparently, for thick samples $> 50$
layers, the self-assembling process loses its effectiveness, as it evident by the increase of
intrinsic disorder and by the cracks  that appear on thick sample
as they start to lift from the substrate.

The physical picture we propose can be checked against consistency
if additional disorder is added to the photonic crystals. This can
be done by doping the original photonic crystal with a controlled
concentration of vacancies.  At a wavelength of 633 nm the values
of $\ell_s$, calculated from the fit to the Lambert-Beer law, are
plotted as a function of vacancy density in Fig. \ref{fig2}b.
The "perfect" crystal (that with 0\% added vacancies)  is highly
ordered as it presents a scattering mean free path of 63 $ \mu$m,
hundreds of times the lattice constant (in this case $a =
0.33$ $\mu$m), and in particular much larger than the Bragg length
(of the thick sample) \cite{JuanPRB} that in our case is $L_B =
(3.8 \pm 0.3)$ $\mu$m. An addition of a very little amount of
defects rapidly decreases the mean free path, hence the quality of
the crystal. In this figure, the inverse scattering mean free path
scales linearly with the vacancy concentration $\rho_v$, as shown
by the black line, which is a fit for $\ell_s ^{-1} = \rho_0
\sigma_0 + \rho_v \sigma_v$ where $\rho_0$ and $\rho_v$ are the
density of intrinsic and intentionally added scatterers and
$\sigma_0$ and $\sigma_v$ their scattering cross-section
respectively.  From the fit of $\ell_s (\rho_v)$ as a function of
the vacancy concentration we can estimate $\sigma_v = (0.016 \pm
0.002)$ $\mu$m$^2$.

   \begin{figure}[!h]
    \begin{center}
   \includegraphics[width=8cm]{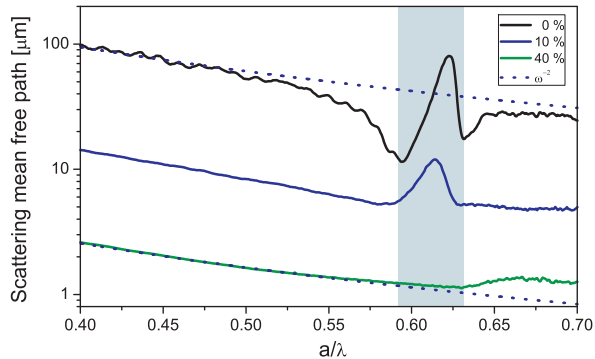}
    \caption{ \label{fig3}
    Figure shows $\ell_s$ as a function of the light wavelength for 0\%, 10\% and 40\% vacancy doped photonic crystals with $d= 237$ nm. The position of the pseudo-gap is shaded
    in cyan. The dotted line shows the
    $\omega^{-2}$
    dependence of$\ell_s$ far from the band-gap.  }
    \end{center}
    \end{figure}

Fig. \ref{fig3} shows the strong chromatic dispersion of
$\ell_s(\omega)$ in the visible range. This  is the signature of
the photonic crystal. In the low energy side of the
pseudo-bandgap, $\ell_s$ takes on a value of the order of $\sim
100$ $\mu$m for sphere diameter d = 237 nm and $\sim$ 500 $\mu$m
for $d = 600$ nm (not shown here), the largest values reported so
far. Previous experiments \cite{Foemius-cbs, Astratov, Vlasov} have
measured $\ell_t$, the transport mean free path, from very thick
($\sim$ 200 $\mu$m) photonic crystals grown by natural
sedimentation \cite{Astratov, Vlasov} or centrifugation
\cite{Foemius-cbs}, and found values in the range 7 $\mu$m $<
\ell_t  < 20 \mu$m.  $\ell_s$ is in general smaller than $\ell_t$
and, therefore, the values found in our experiment represent a
much higher degree of ordering than previous reports.
Far from the band-gap, $\ell_s(\omega)$  varies as $\sim
\omega^{-2}$, dependence that has been
confirmed also in previous experiments \cite{Foemius-cbs} and attributed to
Rayleig-Gans type of scattering.

A simple model for a point-like scatterer in a photonic crystal
can explain the chromatic dispersion of $\ell_s$. We assume that
the scattering is isotropic and, therefore, that the imaginary
part of the green tensor $\mathbb{G}(\mathbf{r},\mathbf{r'})$ can be
approximated  by  $\mathrm{Im}\{\mathbb{G}(\mathbf{r},\mathbf{r'})\} \sim
(\pi c^2/2\omega) \mathcal{D}(\mathbf{r},\omega) \mathbb{I} $. We denote the
$\mathbf{k}$ contribution to the local density of states (or
"projected" density of states) by
$\mathcal{D}_{\mathbf{k}}(\mathbf{r},\omega)$, and calculating the
power radiated by a  dipole we obtain a (spatially averaged)
scattering cross section
        \begin{equation}
        \label{sigma_D}
    \sigma_\mathbf{k}(\omega)\sim F(\omega) \omega^2 \mathcal{D}(\omega)\mathcal{D}_{\mathbf{k}}    (\omega),
    \end{equation}
where $F(\omega)$ is a form factor which takes into account
corrections beyond Rayleigh scattering and $\mathbf{k}$ the wavevector in the incident direction. In our simple model, the
form factor can be replaced by a Rayleigh-Gans factor $F(\omega)
\sim \omega^{-2}$ and the polarizability of the scatterer can be
considered independent of frequency \cite{Foemius-cbs}. Eq.
\ref{sigma_D} states that the scattering cross-section, and hence
the scattering mean free path, has a strong dependence on  the
total $\mathcal{D}(\omega)$ and projected
$\mathcal{D}_{\mathbf{k}}(\omega)$. This dependence typically
disappears in ordinary random media for which the photonic modes
are isotropic and energetically smooth, but is very important for
photonic crystals. At the band-edges of our photonic crystals,
$\ell_s (\omega)$ has a sharp decrease of a factor of up 4 to
$\ell_s (a/\lambda=0.59) = 11 \pm 1$ $\mu$m and then it shoots up
almost an order of magnitude in the band-gap to $\ell_s
(a/\lambda=0.62) = 81 \pm 40$ $\mu$m. Such an 8-fold increase
occurs within just 0.03 in $a/\lambda$ and $\sim$27 nm in
wavelength, around the photonic band-gap. Again, as a comparison,
we show in Fig. \ref{fig3} the frequency dependence of $\ell_s$
for the 10\% and the 40\% vacancy photonic crystal, the latter can
be considered fully disordered. As the vacancy doping is
increased, the profile is smoothed. Firstly the band-edge effect
on $\ell_s$ disappears, as these standing-wave-like states are
very sensitive to disorder. Then, for the 40\% vacancy case, the
effects on $\mathcal{D}(\omega)$ are washed out and the only
feature in $\ell_s$ occurs at the position of the first Mie
resonance of the individual dielectric spheres \cite{PRLmie}. This
weak energy dependence of $\ell_s$ is likely to be the only
residual effect in a very disordered opal, as those grown by
natural sedimentation or centrifugation
\cite{Foemius-cbs, Astratov, Vlasov}, which present superficial
iridescence but are largely bulk-disordered and exhibit standard
light diffusion.

   \begin{figure}[!h]
    \begin{center}
  \includegraphics[width=8.5cm]{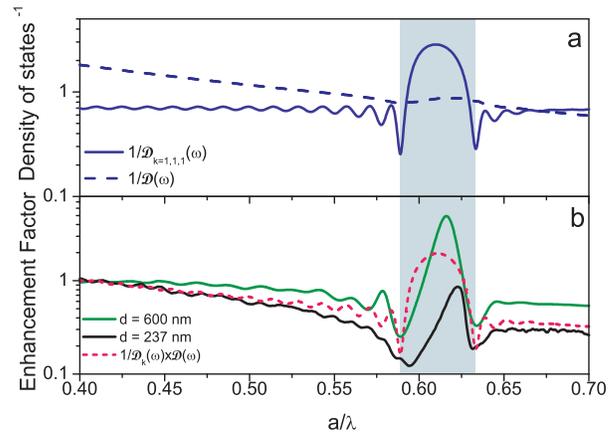}
    \caption{ \label{fig4} (a) Inverse of the total $\mathcal{D}(\omega)$ and the projected
    $\mathcal{D}_{k=(1,1,1)}(\omega)$ along the incident direction
    ($\Gamma-L$ direction) are plotted. (b) Enhancement factor,
    $\ell_s (a/\lambda)$ /$\ell_s (a/\lambda=0.4)$,
    for two opals with no vacancy doping and the product $(\mathcal{D}_k(\omega)-1)$. The position of the pseudo-gap is shaded in cyan. }
    \end{center}
    \end{figure}

Figure \ref{fig4}a shows the inverse of  the total density of
states (dark blue dotted line) that has a very weak modulation at
the gap together with $\mathcal{D}_{\mathbf{k}} (\omega)$ for propagation
parallel to $\mathbf{k} = (111)$ that does have a strong variation at the
gap (violet full line). $\mathcal{D}_{\mathbf{k}} (\omega)$ is calculated
from the inverse of the group velocity \cite{Balian} $v_G(111)$
from ref. \cite{JuanGalli}.

Figure \ref{fig4}b shows the enhancement factor defined as the
ratio of $\ell_s$ to its value far from the gap,
$\ell_s(a/\lambda)/\ell_s(a/\lambda = 0.4 )$, for two different
samples composed by PMMA spheres of 237 nm (black curve) and 600
nm (green curve) in diameter respectively.
The enhancement factor points out the existence of a
photonic pseudo-gap and reveals the variation of the density of
states in the photonic crystal. A clear resonant behavior is
evident. The variation in $\ell_s$ is 8-fold for $d = 237$ nm and
20-fold for $d = 600$ nm, which we attribute to the superior quality
of the lattice.
In figure 4b we plot also
1/$\mathcal{D}(\omega)\mathcal{D}(\omega)_{\mathbf{k}}$ (dashed pink curve)
that, from Eq. \ref{sigma_D}, is expected to reproduce the shape of energy
dependence of $\ell_s(\omega)$. A fair agreement between theory
and experiment is obtained, and the qualitative behavior is well
captured by our simple model.
Although both $\mathcal{D}(\omega)$ and $\mathcal{D}_{\mathbf{k}}(\omega )$
contribute to the strong variation of $\ell_s$, it is evident that
the principal factor responsible for a change in $\ell_s$ (111) is
$\mathcal{D}_{\mathbf{k}}( \omega)$.

Our simple and qualitative model accounts remarkably well for the
shape of the measured $\ell_s(\omega )$ although it does not
account for the asymmetry of its dispersion in the photonic gap.
This effect is related to the available scattering states in other
crystallographic directions close to the incident one
\cite{Cassagne}.
 Our measurement indicates the need for
a more complete theoretical model that should account for all the
modes in all directions with the right scattering probability.

An increase of $\ell_s$ in the band-gap and a decrease at the
band-edge reflects the modified phase space available $\Delta k$
for light scattering when the photonic modes are concentrated
around few $k$-directions or the available scattered states
reduced. This is consistent with John's seminal prediction of a
need for a modified Ioffe-Regel criterion \cite{Saggio} for
scattering in photonic crystals, to include $\Delta k$. In
addition, here we show that as the phase-space is modified, also
$\ell_s$ is altered: light scattering in photonic
crystals is richer than in conventional amorphous media. Complete
photonic band-gap materials, like Si inverted opals, would amplify
the effect here presented and could be proper candidates to
observe Anderson localisation of light.

In conclusion, we show that a controlled smooth transition from
ballistic to diffuse transport in photonic crystals can be induced
by the introduction of extrinsic disorder. We find that
the strength of scattering is closely related to the density of
states, which induces immense, up to 20-fold, variations in the
scattering mean free path.  We propose $\ell_s$ as a robust, easy to measure,
figure of merit in assessing the quality of photonic crystals for
technological applications.
The possibility of controlling light scattering and diffusion in
nano-structured optical media has important implications not only
to test the quality of photonic devices, but also to properly
address the proximity to the onset of Anderson localization in
disordered lattices, or for the spectral control of lasing
emission from disordered/ordered active media \cite{randomlasing}.

We thank J.F. Galisteo-L\'{o}pez for the data of the group
velocity and J. J. S\'{a}enz for fruitful discussion. We are also
indebted to D. Delande, M. Artoni and Ad Lagendijk for invaluable
advice. LSFP acknowledges the finacial suport of the Spanish Ministry of Science and
Innovation through its Juan de la Cierva program. The work was supported by the EU through Network of
Excellence IST-2-511616-NOE (PHOREMOST), CICyT NAN2004-08843-C05,
MAT2006-09062, the Spanish MEC Consolider-QOIT CSD2006-0019 and
the Comunidad de Madrid S-0505/ESP-0200.


\end{document}